\documentclass[runningheads]{llncs}
\usepackage[colorlinks,linkcolor=red,anchorcolor=blue,citecolor=green]{hyperref}
\usepackage{amsmath}
\usepackage{amssymb}
\usepackage{graphicx}

\begin{document}

%\tableofcontents 
%
\title{Differential and integral invariants under \\M\"{o}bius transformation}
\titlerunning{Invariants under M\"{o}bius transformation}

\author{He Zhang\inst{1,2,}\thanks{Corresponding auther.} \and
Hanlin Mo\inst{1,2} \and
You Hao\inst{1,2} \and
Qi Li\inst{1,2} \and
Hua Li\inst{1,2}}
\authorrunning{H. Zhang et al.}
\institute{Key Laboratory of Intelligent Information Processing, Institute of Computing Technology, Chinese Academy of Sciences, China \and
University of Chinese Academy of Sciences, China
\\
\email{zhanghe@ict.ac.cn}}
\maketitle              % typeset the header of the contribution

\begin{abstract}
One of the most challenging problems in the domain of 2-D image or 3-D shape is to handle the non-rigid deformation. From the perspective of transformation groups, the conformal transformation is a key part of the diffeomorphism. According to the Liouville Theorem, an important part of the conformal transformation is the M\"obius transformation, so we focus on M\"obius transformation and propose two differential expressions that are invariable under 2-D and 3-D M\"obius transformation respectively. Next, we analyze the absoluteness and relativity of invariance on them and their components. After that, we propose integral invariants under  M\"obius transformation based on the two differential expressions. Finally, we propose a conjecture about the structure of differential invariants under conformal transformation according to our observation on the composition of above two differential invariants.

\end{abstract}

\section{Introduction}
\vspace{-0.5cm}
\begin{figure}%[H]
%	\centering\includegraphics[scale=0.25]{Flowchart.png}
	\centering\includegraphics[width=\textwidth]{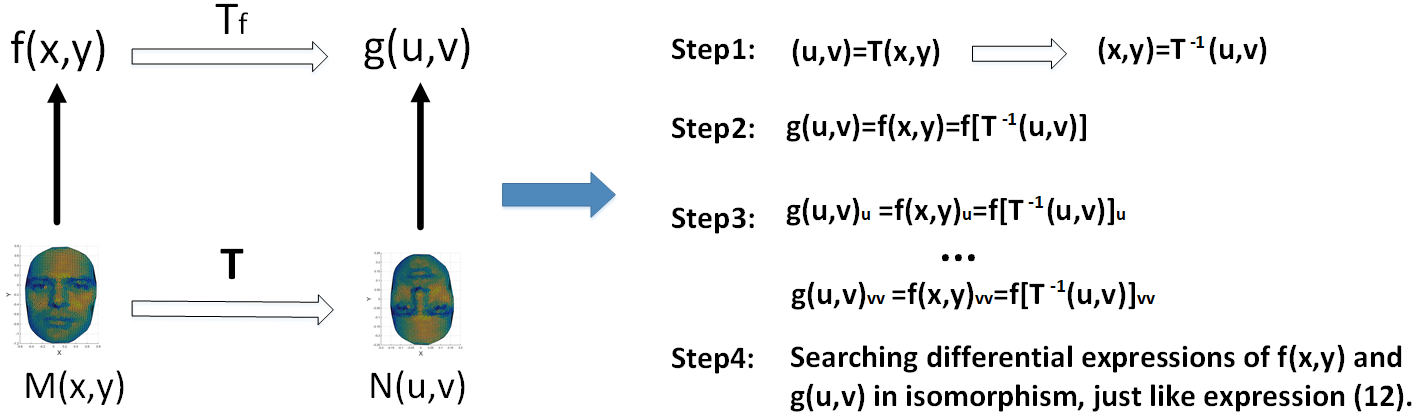}
	\caption{A brief flowchart of the method.}
	\label{fig:one}
\end{figure}
\vspace{-0.5cm}
One of the most challenging problems in the domain of 2-D image or 3-D shape is to handle the non-rigid deformation, especially in the situation of anisotropy, which is universal in the real world. In the viewpoint of transformation groups, the isometric transformation is a prop subgroup of the conformal transformation, which is a prop subgroup of the diffeomorphism. Obviously, the anisotropic non-rigid transformation exceeds the boundary of isometric transformation and contains conformal transformation. Based on the Erlangen program of Klein, geometry is a discipline that studies the properties of space that remain unchanged under a particular group of transformation. In order to solve the anisotropic transformation problem, it is necessary to find the invariants under the conformal transformation.
%\begin{figure}%[H]
%	\centering\includegraphics[scale=0.3]{TransformationClass.png}
%	%	\includegraphics[width=\textwidth]{TransformationClass.png}
%	\caption{The relationships between isometric transformation, conformal transformation and diffeomorphism.}
%	\label{fig:one}
%\end{figure}

%The descriptions of conformal mapping contain angle preservation\cite{Hu2011AClassofIsometric,rustamov2013map,corman2017functional}, metric rescaling\cite{luo2004combinatorial,springborn2008conformal}, preservation of circles\cite{kharevych2006discrete,vaxman2015conformal}, etc. Some key ideas reside in the conformal surface geometry are M\"{o}bius transformations\cite{springborn2008conformal,vaxman2015conformal}, Riemann mapping\cite{gu2003surface,gu2003computing,zeng2011registration,xu2018content}, Cauchy-Riemann equation, Ricci flow\cite{yu2017intrinsic}, etc.

The original motivation of conformal mapping is how to flatten the map of globe, and the Mercator projection produce an angle-preserving map that is very useful for navigation. More generally, the conformal geometry focuses on the shape in which the only measure is angle instead of usually length. The descriptions of conformal mapping contain angle preservation\cite{Hu2011AClassofIsometric,rustamov2013map,corman2017functional}, metric rescaling\cite{luo2004combinatorial,springborn2008conformal}, preservation of circles\cite{kharevych2006discrete,vaxman2015conformal}, etc. Some key ideas reside in the conformal surface geometry are Dirac equation\cite{crane2011spin}, Cauchy-Riemann equation\cite{mullen2008spectral}, M\"{o}bius transformations\cite{springborn2008conformal,vaxman2015conformal}, Riemann mapping\cite{gu2003surface,gu2003computing,zeng2011registration,xu2018content}, Ricci flow\cite{yu2017intrinsic}, etc. The conformal geometry lies between the topology geometry and the Riemannian geometry, it studies the invariants of the conformal transformation group. The \emph{conformal structures}\cite{gu2003surface,gu2003computing} based on the theories of Riemann surfaces are invariants under conformal transformation. According to conformal geometry\cite{farkas1992riemann}, the \emph{shape factor}\cite{gu2003surface} and \emph{conformal module}\cite{zeng2011registration} are conformal invariants. Moreover, the \emph{conformal inner product}\cite{rustamov2013map} defined by an inner product of function is also changeless under conformal transformation. According to the Liouville Theorem\cite{liouville1850extension,gehring1992topics}, the M\"{o}bius transformation plays an important role in conformal mapping.

The definition of M\"{o}bius transformation\cite{rasila2006introduction} shows that it is compounded by a series of simple transformations: Translation, Stretching, Rotation, Reflection and Inversion.
In the domain of invariants under translation, stretching and rotation transformations, the Geometric moment invariants(GMIs)\cite{xu2008geometric} and the ShapeDNA\cite{li2017shape} show a general method to generate the moment invariants; Hu et al\cite{hu2009construction} proposed a general construction method of surface isometric moment invariants based on the intrinsic metric. In the domain of invariants under reflection transformation, the chiral invariants\cite{zhang2017fast} show the moment invariants based on the generating functions of ShapeDNA\cite{li2017shape}. In the domain of invariants under conformal transformation, Hu\cite{Hu2011AClassofIsometric} proposed limited conformal invariants based on geodesic tangent vectors. In the domain of invariants under M\"{o}bius transformation, the expression $(H^{2}-K)dA$ proposed by Blaschke\cite{biaschke1929vorlesungen} is proved to be a conformal invariant by Chen\cite{chen1973invariant}; based on the \emph{Gauss-Bonnet} Theorem, White\cite{white1973global} proposed that \emph{$\int_{M}H^{2}dA$} is a global conformal invariant if $M$ is an oriented and closed surface. The \emph{Gauss-Bonnet} Theorem associates the differential expression(Gaussian curvature) of the surface $S$ with its topological invariant $\chi(S)$(the Euler's characteristic). This great theorem motivates us to explore the differential invariants under the M\"{o}bius transformation since the differential expressions play essential roles in some procedures of physics, mathematics, computer science and other fields. In the domain of differential invariants, rotation and affine differential invariants were proposed by Olver\cite{olver1995equivalence} based on the moving frame method; a special type of affine differential invariants was presented by Wang et al\cite{wang2013affine}; Li et al\cite{li2018image} prove the existence of projective moment invariants of images with relative projective differential invariants; the research\cite{li2017isomorphism} on the relationship between differential invariants and moment invariants show that they are isomorphic under affine transformation. 

In this article, we study invariants by combining functional map\cite{ovsjanikov2012functional} and the derivatives of function(see Fig.\ref{fig:one}). In section 2, we show the background of this paper. In section 3, we propose the invariants under M\"obius transformation. In section 4, we show another M\"obius invariant from the functional view. Finally, we propose a conjecture about the structure of differential invariants under conformal transformation. The main contributions of this paper are as follows.
\begin{itemize}
	\item We propose two differential expressions that are invariant under 2-D and 3-D M\"{o}bius transformation respectively. According to the Liouville Theorem, the 3-D differential invariant is a conformal invariant.
	\item Based on the analysis on absoluteness and relativity of invariance about the two differential expressions and their components, we propose integral invariants under  M\"obius transformation.
	\item We propose a conjecture about the composition of differential invariants under conformal transformation.
\end{itemize}

% Head 1

\section{Notion and Background}
\subsection{Notion}
The formulation in this paper is same with the functional maps framwork\cite{ovsjanikov2012functional}. Assuming $M$ and $N$ are two manifolds, a bijective mapping $T:M \rightarrow N$ induces the transformation $T_{F}:\mathcal{F}(M,\mathbb{R})\rightarrow\mathcal{F}(N,\mathbb{R})$ of derived quantities, where $\mathcal{F}(\cdot,\mathbb{R})$ is scalar function defined on manifold. It means that any function $f:M\rightarrow\mathbb{R}$ have a counterpart function $g:N\rightarrow\mathbb{R}$ and $g=f\circ T^{-1}$.

To make the invariants under M\"{o}bius transformation clear, we partially modify original definition and theorem in this paper with this formulation.
\subsection{Theoretic Background}
\label{Background}

According to the Liouville Theorem\cite{liouville1850extension}, the only conformal mapping in $R^{n}(n>2)$ are M\"{o}bius transformation\cite{haantjes1937conformal,rasila2006introduction,kuhnel2007liouville}. Furthermore, the Generalized Liouville Theorem shows that any conformal mapping defined on $D$($D\in \overline{\mathbb{R}}^{n}, n>2$) must be a restriction of M\"{o}bius transformation.
\begin{theorem}[Generalized Liouville Theorem\cite{gehring1992topics}]\label{LiouvilletTheorem}
	Suppose that $D$, $D'$ are domains in $\overline{\mathbb{R}}^n$ and that $T:D \rightarrow D'$ is a homeomorphism. If $n=2$, then $T$ is 1-quasiconformal if and only if $T$ or its complex conjucate is a meromorphic function of a complex variable in $D$. If $n\geq 3$, then $T$ is 1-quasiconformal if and only if $T$ is the restriction to $D$ of a M\"{o}bius transformation, i.e., the composition of a finite number of reflections in $(n-1)$-spheres and planes.
\end{theorem}

%Now we know that the M\"{o}bius transformation plays an important role in conformal mapping.

Next, we will show the common expressions of M\"{o}bius transformation in different dimensions($n\geq2$).

In the filed of complex analysis, a M\"{o}bius transformation could be expressed as
% Numbered Equation
\begin{equation}
\label{eqn:01}
T(z)=\dfrac{az+b}{cz+d},
\end{equation}
where $a, b, c, d, z\in \mathbb{C}$, $ad-bc\neq 0$. Based on the Liouville Theorem\cite{liouville1850extension}, every M\"{o}bius transformation in higher dimensions could be given with the form
% Numbered Equation
\begin{equation}
\label{eqn:02}
T(x)=b+\dfrac{\gamma A (x-a)}{\parallel x-a\parallel_{2}^{\epsilon}},
\end{equation}
where $x, a, b \in \mathbb{R}^{n}$, $\epsilon$ is 0 or 2, $\gamma \in \mathbb{R}$ and $A \in \mathbb{R}_{n \times n}$ is an orthogonal matrix. The choice of $\epsilon$ decides if $T(x)$ contains inversion transformation, and the sign of $det(A)$ decides if $T(x)$ contains reflection transformation.

More generally, a M\"{o}bius transformation could be composed of a series of simple transformations, the definition of M\"{o}bius transformation is as below. 
%\vspace{-0.9cm} 
\begin{figure}
	\centering
	\includegraphics[width=\textwidth]{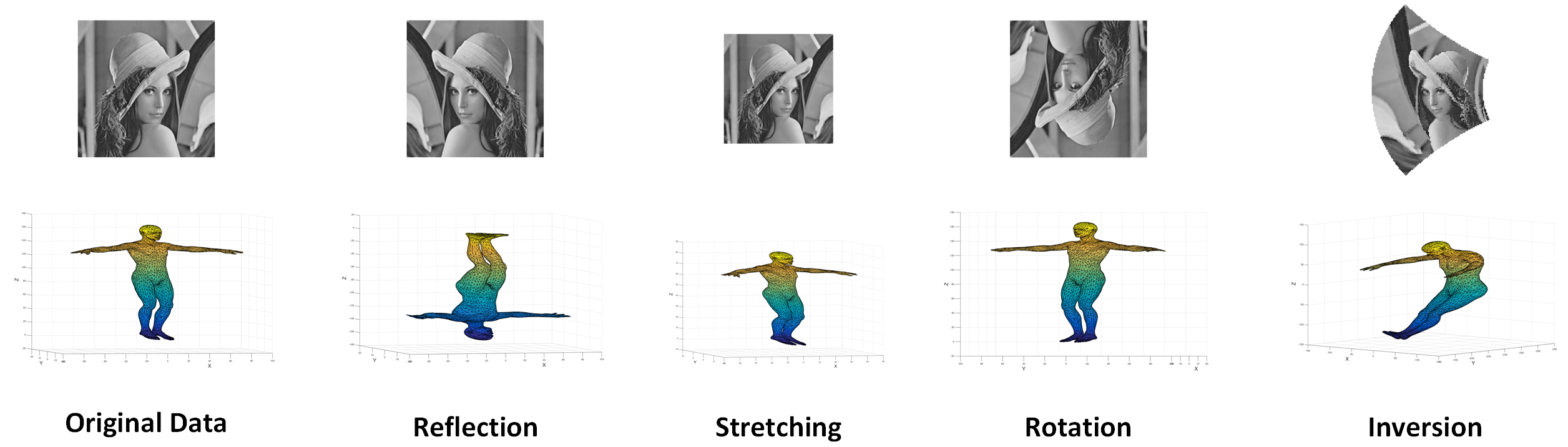}
	\caption{Some elementary transformations of M\"obius transformation.}
	\label{fig:two}
\end{figure}
%\vspace{-0.9cm} 
\begin{definition}[M\"{o}bius transformation\cite{rasila2006introduction}\label{def:01}]A n-dimension M\"{o}bius transformation is a homemorphism of $\overline{\mathbb{R}}^n$(the one-point compactification of $\mathbb{R}^n$), it is a mapping $T:\overline{\mathbb{R}}^n\rightarrow\overline{\mathbb{R}}^n$ that is a finite composition of the following elementary transformations($x\in\mathbb{R}^n$):
	
	(1)Translation: $T_{a}(x)=x+a$, $a\in\mathbb{R}^n$.
	
	(2)Stretching: $S_{s}(x)=sx$, $s\in\mathbb{R}$ and $s>0$.
	
	(3)Rotation: $Rot_{R}(x)=Rx$, $R\in\mathbb{R}_{n\times n}$ and $R$ is an orthogonal matrix.
	
	(4)Reflection about plane $P(a,t)$: $Ref_{a,t}(x)=x-2(a^{T}x-t)a$, $a\in\mathbb{R}^n$ is the normal vector of $P(a,t)$, $t\in\mathbb{R}$ is the distance from the origin to $P(a,t)$.
	
	(5)Inversion about sphere $S^{n-1}(a,r)$: $I_{a,r}(x)=a+\dfrac{r^{2}(x-a)}{\parallel x-a\parallel_{2}^{2}}$,  $a\in\mathbb{R}^n$ is the inversion center, $r$ is the inversion radius.
\end{definition}

\section{M\"{o}bius Invariants}
% Head 2
\subsection{Inversion Invariants}\label{BasicMethod}
In order to derive the differential invariant under inversion transformation $I_{a,r}$, in the 2-D situation we assume that the $T_{I_{a,r}}$ map the function $f(x,y)$ on domain $D\subset \overline{\mathbb{R}}^n$ to $g(u,v)$ on domain $D'\subset \overline{\mathbb{R}}^n$, where $(u,v)=I_{a,r}(x,y)$ and $g(u,v)=f(x,y)$, this means that the coordinates transformations under $I_{a,r}$ are as follows. 
\begin{equation}
\label{eqn:03}
u=a_{x}+\dfrac{r^{2}(x-a_{x})}{(x-a_{x})^{2}+(y-a_{y})^{2}}
\end{equation}
\begin{equation}
\label{eqn:04}
v=a_{y}+\dfrac{r^{2}(y-a_{y})}{(x-a_{x})^{2}+(y-a_{y})^{2}}
\end{equation}
At the same time, it means that the coordinates transformations under $I_{a,r}^{-1}$ are as follows.
\begin{equation}
\label{eqn:05}
x=a_{x}+\dfrac{r^{2}(u-a_{x})}{(u-a_{x})^{2}+(v-a_{y})^{2}}
\end{equation}
\begin{equation}
\label{eqn:06}
y=a_{y}+\dfrac{r^{2}(v-a_{y})}{(u-a_{x})^{2}+(v-a_{y})^{2}}
\end{equation}

Based on $g(u,v)=f(x,y)$ and the equations (\ref{eqn:05})(\ref{eqn:06}), we obtain the relationships between the partial derivatives of $g(u,v)$ and $f(x,y)$ as follows.
\begin{eqnarray}
g_{u}&=&f_{x}x_{u}+f_{y}y_{u}\\
g_{v}&=&f_{x}x_{v}+f_{y}y_{v}\\
g_{uu}&=&(f_{xx}x_{u}+f_{xy}y_{u})x_{u}+f_{x}x_{uu}+(f_{yx}x_{u}+f_{yy}y_{u})y_{u}+f_{y}y_{uu}\\
g_{uv}&=&(f_{xx}x_{v}+f_{xy}y_{v})x_{u}+f_{x}x_{uv}+(f_{yx}x_{v}+f_{yy}y_{v})y_{u}+f_{y}y_{uv}\\
g_{vv}&=&(f_{xx}x_{v}+f_{xy}y_{v})x_{v}+f_{x}x_{vv}+(f_{yx}x_{v}+f_{yy}y_{v})y_{v}+f_{y}y_{vv}
\end{eqnarray}
Then we obtain a 2-D equation under the inversion transformation, it is
\begin{equation}
\label{equ:2D-INV}
\dfrac{g_{uu}+g_{vv}}{g_{u}^{2}+g_{v}^{2}}=\dfrac{f_{xx}+f_{yy}}{f_{x}^{2}+f_{y}^{2}}
\end{equation}
This means that 
\begin{equation}
\label{equ:2D-MI}
\dfrac{f_{xx}+f_{yy}}{f_{x}^{2}+f_{y}^{2}}
\end{equation}
is a differential invariant under inversion transformation. We use the same method in 3-D situation and obtain a differential invariant under the inversion transformation, it is
\begin{equation}
\label{equ:3D-CI}
\dfrac{f_{A}+f_{B}}{(f_{x}^{2}+f_{y}^{2}+f_{z}^{2})^{2}}
\end{equation}
where
\begin{equation}
\begin{aligned}
&f_{A}=(f_{xx}+f{yy}+f{zz})(f_{x}^{2}+f_{y}^{2}+f_{z}^{2})\\
&f_{B}=f_{x}^{2}f_{xx}+f_{y}^{2}f_{yy}+f_{z}^{2}f_{zz}+2f_{x}f_{xy}f_{y}+2f_{x}f_{xz}f_{z}+2f_{y}f_{yz}f_{z}
\end{aligned}
\end{equation}

\subsection{The Boundary of Invariance}\label{BoundaryofInvariance}
We have shown that (\ref{equ:2D-MI}) and (\ref{equ:3D-CI}) are differential invariants under inversion transformation. It is obvious that they are invariants under translation transformation. We prove that (\ref{equ:2D-MI}) and (\ref{equ:3D-CI}) are also differential invariants under rotation, stretching and reflection transformations(see \href{https://github.com/duduhe/Differential-and-integral-invariants-under-Mobius-transformation/blob/master/Appendix.pdf}{Appendix A} for a proof). According to the definition of M\"{o}bius transformation, we conclude that the differential expression (\ref{equ:2D-MI}) is a differential invariant under 2-D M\"obius transformation. Furthermore, with the Generalized Liouville Theorem we obtain that (\ref{equ:3D-CI}) is a conformal invariant.

\subsection{Absoluteness and Relativity of Invariance}
If expression $Inv_{T}$ is an invariant under transformation $T$, the transformed expression $Inv_{T}^{'}$ satisfies 
\begin{equation}
Inv_{T}^{'}=W_{T}\cdot Inv_{T}
\end{equation}
where $W_{T}$ is an expression related to $T$. In this context, $Inv_{T}$ is an absolute invariant if $W_{T}\equiv 1$, otherwise, $Inv_{T}$ is a relative invariant. Base on the analysis in \ref{BoundaryofInvariance}, (\ref{equ:2D-MI}) is an absolute invariant under M\"obius transformation and (\ref{equ:3D-CI}) is an absolute invariant under conformal transformation. Next, we will show the numerator and denominator of (\ref{equ:2D-MI}) or (\ref{equ:3D-CI}) are relative invariants. 

In the derivation of 2-D inversion invariants, we obtain that $W_{I_{a,r}}=||J||^{-1}$ for the numerator and denominator of (\ref{equ:2D-MI}), this means
\begin{equation}
\label{RI-2D-1}
g_{uu}+g_{vv}=||J||^{-1}(f_{xx}+f_{yy})
\end{equation}
\begin{equation}
\label{RI-2D-2}
g_{u}^{2}+g_{v}^{2}=||J||^{-1}(f_{x}^{2}+f_{y}^{2})
\end{equation}
%are relative differential invariants under 2-D inversion transformation,
where $|J|$ is the determinant of Jacobian matrix of transformation $I_{a,r}$, $||J||$ is the absolute valve of $|J|$. In 3-D situation, we obtain $W_{I_{a,r}}=||J||^{-\frac{4}{3}}$ for the numerator and denominator of (\ref{equ:3D-CI}). In the stretching transformation, we obtain $W_{S}=||J||^{-1}$ in 2-D situation, and $W_{S}=||J||^{-\frac{4}{3}}$ in 3-D situation. We also obtain that $W_{T}=1$ for the numerator and denominator of (\ref{equ:2D-MI}) or (\ref{equ:3D-CI}) under translation, rotation and reflection transformations. 

The result of absoluteness and relativity of invariance on (\ref{equ:2D-MI}) and (\ref{equ:3D-CI}) is shown in Table \ref{tab:one}.
%\vspace{-0.5cm} 
\begin{table}%
	\caption{The form of $W_{T}$ under Transformations}
	\label{tab:one}
	\begin{minipage}{\columnwidth}
		\begin{center}
			\begin{tabular}{|c|c|c|c|c|c|}
				\hline
				Expression& Translation&Stretching &Rotation &Reflection &Inversion \\
				\hline
				(\ref{equ:2D-MI}) and (\ref{equ:3D-CI})    & 1 & 1&1 & 1&1\\
				Num\footnote{Num means the numerator of fraction .}/Den of (\ref{equ:2D-MI})   & 1&$||J||^{-1}$ &1 &1 &$||J||^{-1}$  \\
				Num/Den\footnote{Den means the denominator of fraction.} of (\ref{equ:3D-CI})   & 1&$||J||^{-\frac{4}{3}}$ &1 &1 &$||J||^{-\frac{4}{3}}$ \\
				\hline
			\end{tabular}
		\end{center}
		%		\bigskip
		\centering		
		%		\footnotesize\emph{Source:} This is a table
		%		sourcenote. This is a table sourcenote. This is a table
		%		sourcenote.
		%		\emph{Note:} This is a table footnote.
	\end{minipage}
\end{table}%
%\vspace{-0.8cm} 
\subsection{Multiscale and Quantity}
\label{Integral Invariants}
Assuming $f(x,y)$ is a regular parameter surface $S$ defined on $D$, if $T_{F}$ transform $f(x,y)$ defined on $D$ to $g(u,v)$ defined on $D'$ and $g(u,v)=f(x,y)$, based on the change of variable theorem\cite{lax1999change} for multiple integrals and Table \ref{tab:one} we obtain that
\begin{equation}
\label{INT_2D_1}
\iint_{D'}(g_{uu}+g_{vv})dudv=\iint_{D}W_{T}(f_{xx}+f_{yy})||J_{T}||dxdy=\iint_{D}(f_{xx}+f_{yy})dxdy
\end{equation}
\vspace{-0.5cm} 
\begin{equation}
\label{INT_2D_2}
\iint_{D'}(g_{u}^{2}+g_{v}^{2})dudv=\iint_{D}W_{T}(f_{x}^{2}+f_{y}^{2})||J_{T}||dxdy=\iint_{D}(f_{x}^{2}+f_{y}^{2})dxdy
\end{equation}
where $||J_{T}||$ is the area extension factor, so we obtain that 
\begin{equation}
\iint_{D}(f_{xx}+f_{yy})dxdy
\end{equation}
\begin{equation}
\iint_{D}(f_{x}^{2}+f_{y}^{2})dxdy
\end{equation}
are integral invariants under 2-D M\"obius transformation. In the same way, we obtain that
\begin{equation}
\iiint_{D}(f_{x}^{2}+f_{y}^{2}+f_{z}^{2})^{\frac{3}{2}}dxdydz
\end{equation}
\begin{equation}
\iiint_{D}(f_{A}+f_{B})^{\frac{3}{4}}dxdydz
\end{equation}
are integral invariants under 3-D conformal transformation. 

Actually a differential expression  $Inv_{T}$ of function $f$ defined on domain $D_{f}$ accurately characterize $f$ at point of $D_{f}$, it provides extremely wide space to describe the function $f$.

\paragraph{\textbf{Multiscale of Invariants}} Assuming $F_{i}(Inv_{T})$ is a function of $Inv_{T}$, a general method to construct descriptors in different scale is the integral of $\int_{D_{j}}F_{i}(Inv_{f})dA$ on region $D_{j}$($D_{j}\subset D_{f}$) with different size, and when $D_{j}=D_{f}$ the result is a global invariant, for example, the Willmore energy $\int(H^2-K)dA$\cite{biaschke1929vorlesungen} applied in the theory of surfaces\cite{willmore2000surfaces}, digital geometry processing\cite{bobenko2005discrete} and other fields. 

In this view, the only difference between invariant with specify-scale and global invariant is the definition domain, the construction method of specify-scale invariant is same with global invariant. The former could be elaborately modified by selecting domain of integration in different applications.

\paragraph{\textbf{Quantity of Invariants}} A general method to construct a large number of invariants is using various functions $F_{i}(Inv_{T})$ with these functions are independent of each other\cite{brown1935functional}. We just show a simple method to construct integral invariants based on differential invariants and integral, in addition, more invariant forms can be constructed with differential invariants. Next, we give a possible form of invariants under M\"obius transformation: 
\begin{equation}
\label{GeneralMI-2D-1}
\iint_{D}\dfrac{(f_{xx}+f_{yy})^{n+1}}{(f_{x}^{2}+f_{y}^{2})^n}dxdy
\end{equation}
\begin{equation}
\label{GeneralMI-2D-2}
\iint_{D}\dfrac{(f_{x}^{2}+f_{y}^{2})^{n+1}}{(f_{xx}+f_{yy})^n}dxdy 
\end{equation}
\begin{equation}
\label{GeneralCI-3D-1}
\iiint_{D}\dfrac{(f_{A}+f_{B})^{\frac{3}{4}(n+1)}}{(f_{x}^{2}+f_{y}^{2}+f_{z}^{2})^{\frac{3}{2}n}}dxdydz
\end{equation}
\begin{equation}
\label{GeneralCI-3D-2}
\iiint_{D}\dfrac{(f_{x}^{2}+f_{y}^{2}+f_{z}^{2})^{\frac{3}{2}(n+1)}}{(f_{A}+f_{B})^{\frac{3}{4}n}}dxdydz
\end{equation}
if the denominators of (\ref{GeneralMI-2D-1}), (\ref{GeneralMI-2D-2}), (\ref{GeneralCI-3D-1}), (\ref{GeneralCI-3D-2}) are not zero.

\subsection{Another Conformal Invariant}\label{OtherInvarian_1}
The expression $(H^2-K)dA$ proposed by Biacchke\cite{biaschke1929vorlesungen} has been proved to be an invariant under M\"{o}bius transformation\cite{chen1973invariant,white1973global}. 
It differs from our method in two important respects: the domain of transformation and the number of functions participated in invariants(see detailed expression at \href{https://github.com/duduhe/Differential-and-integral-invariants-under-Mobius-transformation/blob/master/Appendix.pdf}{Appendix B}).
%The method of this paper focuss on the only function $f(x,y)$ on definition domain $D_{f}$ and the transformation is about the $D_{f}$, the invariants are composed by differential expressions of $f(x,y)$. The $(H^2-K)dA$ of $r(u,v)=(x(u,v),y(u,v),z(u,v))$ focuss on three functions $x(u,v)$, $y(u,v)$, $z(u,v)$ defined on $D_{r}$, and the transformation is about $x$, $y$ and $z$. In the perspective of differential geometry, $dA=\sqrt{EG-F^2}dudv$, we obtain that
%\begin{equation}
%(H^2-K)dA=(H^2-K)\sqrt{EG-F^2}dudv.
%\end{equation}
%The key observation is that M\"{o}bius transformation occurs in the space of $r(u,v)=(x(u,v),y(u,v),z(u,v))$ rather than  $D_{r}$, so we obtain that 
%\begin{equation}
%\label{H2_KsqrtEG_F2}
%(H^2-K)\sqrt{EG-F^2}
%\end{equation}
%is a differential invariant composed by differential expressions of $x(u,v)$, $y(u,v)$ and $z(u,v)$ (see detailed expression at \href{https://github.com/duduhe/Differential-and-integral-invariants-under-Mobius-transformation/blob/master/Appendix.pdf}{Appendix B}), and the integral expression
%\begin{equation}
%\iint_{D_{r}}(H^2-K)\sqrt{EG-F^2}dudv
%\end{equation}
%is a global invariant under M\"{o}bius transformation\cite{chen1973invariant,white1973global}.

\section{Conjecture of Conformal Invariants}

We have shown that (\ref{equ:2D-MI}) is a M\"obius invariant and (\ref{equ:3D-CI}) is a conformal invariant. However, the fascinating part of (\ref{equ:2D-MI}) or (\ref{equ:3D-CI}) is that the differential expressions
\begin{equation}
%\label{DRI-1}
f_{x}^{2}+f_{y}^{2}\quad or\quad f_{x}^{2}+f_{y}^{2}+f_{z}^{2}
\end{equation}
\begin{equation}
%\label{DRI-2}
f_{xx}+f_{yy}\quad or\quad f_{xx}+f_{yy}+f_{zz}
\end{equation}
\begin{equation}
%\label{DRI-3}
f_{x}^{2}f_{xx}+f_{y}^{2}f_{yy}+f_{z}^{2}f_{zz}+2f_{x}f_{xy}f_{y}+2f_{x}f_{xz}f_{z}+2f_{y}f_{yz}f_{z}
\end{equation}
are differential invariants under rigid transformation. Based on this observation and the fact that the differential expressions play important roles in transformation, we have a bold conjecture about the structure of differential invariants under conformal transformation.

\paragraph{\textbf{Conjecture:}}\emph{The differential invariants under conformal transformation are composed of differential invariants under rigid transformation in a self-consistent manner.}

One of the possible self-consistent forms in n-dimensional Euclidean space may be
\begin{equation}
\sum_{i=1}^{n-1}\dfrac{\prod_{j=1}^{a_{i}}DRI_{j}}{(f_{x_{1}}^2+f_{x_{2}}^2+\cdots+f_{x_{n}}^2)^{n-1}}
\end{equation}
where $DRI$ is differential invariant under rigid transformation. 

\section{Experimental Results}
We choose a human face model from TOSCA database and treat the z-coordinate value of vertexes of the triangle mesh as a function $f$ defined on x-coordinate and y-coordinate, i.e. $z=f(x,y)$.  With least square method, the coordinates of a vertice and its 1-ring neighbors were used to estimate parameters in Taylor expansion of $f$ at the vertice; in order to guarantee the accuracy of descriptor calculation, we only consider vertexes that are located inside the mesh and have enough 1-ring neighbors. After that, we calculate a descriptor at the vertice and the descriptor is composed by (\ref{equ:2D-MI}), (\ref{GeneralMI-2D-1}) and (\ref{GeneralMI-2D-2}) with different $n(\geq0)$. Moreover, in integral invariants, the area $A_{vert}$ around a vertice is determined by Mixed Voronoi cell. 

We deform the definition domain of $f$ with reflection, stretching, rotation and inversion transformation(Fig.\ref{fig:three}). In reflection transformation, $a=(1,0)$ and $t=0$; the $s$ in stretching transformation is 2; in rotation transformation the original data is rotated 90 degrees counterclockwise; in inversion transformation the inversion center is $(0,1000)$ and inversion radius is 500(see more explanation about experiments at \href{https://github.com/duduhe/Differential-and-integral-invariants-under-Mobius-transformation/blob/master/Appendix.pdf}{Appendix C}).

\subsection{Stability of Invariants}
\vspace{-0.5cm} 
\begin{figure}
	\centering
	\includegraphics[scale=0.17]{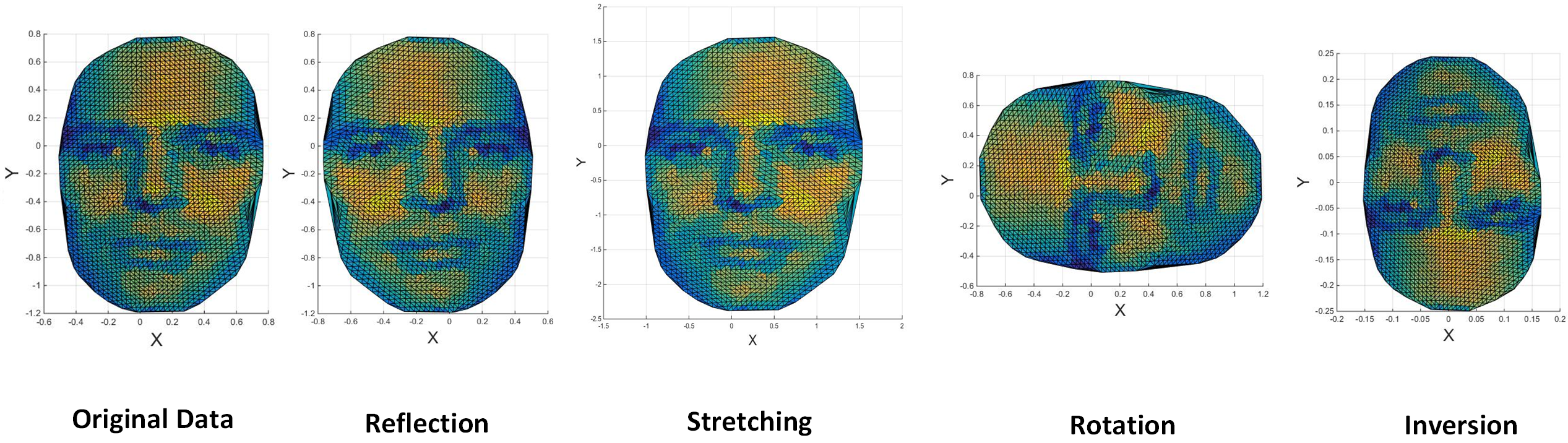}
	\caption{Elementary transformations of M\"obius transformation on human face model.}
	\label{fig:three}	
\end{figure}
\vspace{-0.5cm} 
In this experiment we choose $n=0,\ 1$ and the integral invariants is calculated at the local area of each vertex. After we obtain a 5-dimension descriptor at vertexes of the five mesh in Fig.\ref{fig:three}, we calculate the average error of each dimension of the descriptor. In addition, we choose an isometric invariant at the vertex, the Laplacian operator, to compare with above invariants. The average error of each dimension is calculated by the following formula
\begin{equation}
Err=\frac{1}{N}\sum_{i}\frac{|Inv_{T;i}-Inv_{O;i}|}{|Inv_{T;i}|+|Inv_{O;i}|}\times100\%
\end{equation}
where $Inv_{O;i}$ is the value of invariant at vertex $i$ on original data, $Inv_{T;i}$ is the value of invariant at vertex $i$ on deformed data, and $N$ is the total number of vertexes participated in the calculation. The result of this experiment is in Table \ref{tab:two}, it shows that (\ref{equ:2D-MI}) ,(\ref{GeneralMI-2D-1}) and (\ref{GeneralMI-2D-2}) are invariants under M\"obius transformations.
% Table
%\setlength{\belowcaptionskip}{-10.cm}
%\vspace{-0.5cm} 
\begin{table}
	\caption{The average error of Laplacian operator and M\"obius invariants.}
	\label{tab:two}
	\begin{minipage}{\columnwidth}
		\begin{center}
			\begin{tabular}{|c|c|c|c|c|}
				\hline\centering	
				Expression &Reflection&Stretching &Rotation&Inversion \\
				\hline
				$f_{xx}+f_{yy}$    & $0$&$6.00\times10^{1}$ &$4.82\times10^{-13}$&$8.82\times10^{1}$\\
				\hline
				$\frac{f_{xx}+f_{yy}}{f_{x}^{2}+f_{y}^{2}}$   &$0$&$1.20\times10^{-12}$ &$1.33\times10^{-12}$ &$1.98\times10^{-3}$ \\
				\hline
				$\iint_{D}(f_{xx}+f_{yy})dxdy$ &$0$&$4.38\times10^{-13}$ &$4.82\times10^{-13}$ &$1.69\times10^{-1}$ \\
				\hline
				$\iint_{D}(f_{x}^{2}+f_{y}^{2})dxdy$ &$0$&$1.21\times10^{-12}$ &$1.27\times10^{-12}$ &$1.69\times10^{-1}$ \\
				\hline
				$\iint_{D}\frac{(f_{xx}+f_{yy})^{2}}{f_{x}^{2}+f_{y}^{2}}dxdy$   &$0$&$1.24\times10^{-12}$ &$1.47\times10^{-12}$ &$1.70\times10^{-1}$ \\
				\hline
				$\iint_{D}\frac{(f_{x}^{2}+f_{y}^{2})^{2}}{f_{xx}+f_{yy}}dxdy$   &$0$&$2.39\times10^{-12}$ &$2.58\times10^{-12}$ &$1.70\times10^{-1}$ \\
				\hline
			\end{tabular}
		\end{center}
		\centering		
	\end{minipage}
\end{table}%
%\vspace{-0.9cm} 
\subsection{Discrimination of Invariants}
In this experiment  we use the 5-dimension descriptor of vertex at original to match its corresponding vertex in the deformed mesh with nearest neighbor rule, the metric between vertexes is standardized Euclidean distance. The error rate (percentage) of this experiment is in Table \ref{tab:three}.
%\vspace{-0.5cm} 
\begin{table}%
	\caption{The error rate (percentage) of M\"obius invariants in vertex matching.}
	\label{tab:three}
	\begin{minipage}{\columnwidth}
		\begin{center}
			\begin{tabular}{|c|c|c|c|}
				\hline
				Reflection&Stretching &Rotation&Inversion \\
				\hline
				$0$&$0$ &$0$&$0.87$\\
				\hline
			\end{tabular}
		\end{center}
		\centering		
	\end{minipage}
\end{table}%
\vspace{-0.8cm} 
%This experiment shows the potential of M\"{o}bius invariants in matching task under the conformal deformation scenario. Fig.\ref{fig:four} shows some matching-fail situations, where the white point is the real vertex and the red point is the matching vertex.
%The reason for most matching failures is that the white vertex and red vertex have very similar functional distribution environments.
\begin{figure}
	\centering
	\includegraphics[scale=0.21]{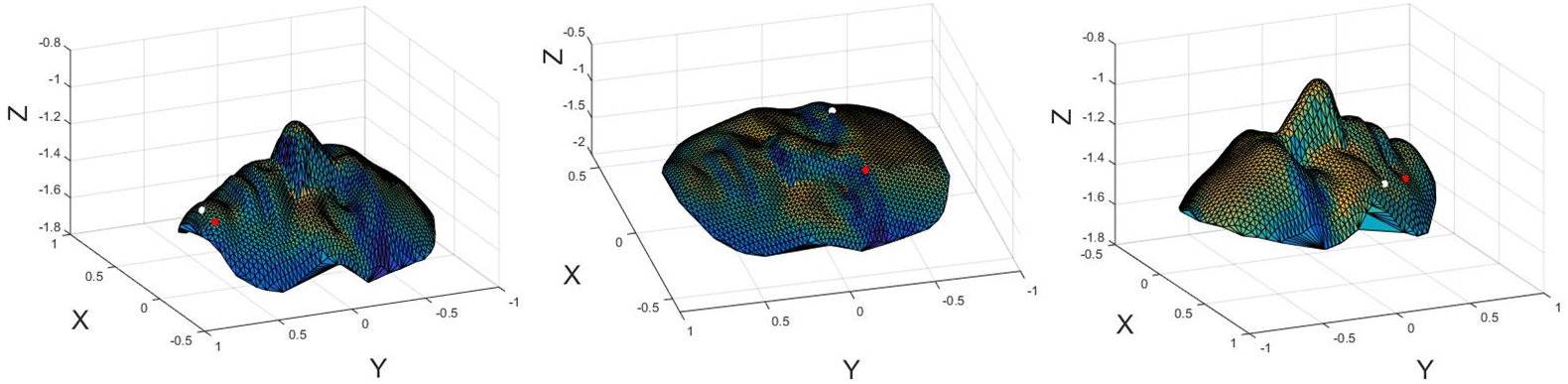}
	\caption{Some situations where vertex matching fails.}
	\label{fig:four}	
\end{figure}

In conformal deformation scenario, this experiment shows the potential of M\"{o}bius invariants in matching task.  Fig.\ref{fig:four} shows some matching-fail situations, where the white point is the real position and the red point is the matching vertex. The reason for most matching failures is that the original white vertex and deformed red vertex have similar functional distribution environments. 
\section{Conclusions}
In this article, we propose two differential invariants under  2-D and 3-D M\"{o}bius transformation respectively, in particular, the 3-D expression is a conformal invariant according to the Liouville Theorem. After that, we analyze the absoluteness and relativity of invariance on the two expressions and their components, and we show an integral construction method that targets to the multiscale and quantity of invariant, the experimental results show that the invariants proposed in this paper perform well. Furthermore, we show another M\"obius invariant from the functional view. Finally, we propose a conjecture about the structure of differential invariants under conformal transformation.

This article shows a method of combining functional map and derivatives of function to study conformal invariant, more research about the differential invariants under conformal transformation is necessary in the future. In addition to practical application solutions based on M\"{o}bius invariants, questing the generative structure of conformal differential invariant is also an interesting topic.

\section{Acknowledgment}
	
The authors would like to thank Dr. Antti Rrasila of Aalto University for providing help on how to distinguish M\"obius invariants and conformal invariants.

This work was partly funded by National Key R\&D Program of China (No. 2017YFB1002703) and National Natural Science Foundation of China (Grant No.60873164, 61227802 and 61379082).

% Bibliography
\bibliographystyle{splncs04}
\bibliography{sample-bibliography}

\begin{thebibliography}{10}
\providecommand{\url}[1]{\texttt{#1}}
\providecommand{\urlprefix}{URL }
\providecommand{\doi}[1]{https://doi.org/#1}

\bibitem{biaschke1929vorlesungen}
Biaschke, W.: Vorlesungen {\"u}ber differentialgeometrie iii (1929)

\bibitem{bobenko2005discrete}
Bobenko, A.I., Schr{\"o}der, P.: Discrete willmore flow  (2005)

\bibitem{brown1935functional}
Brown, A.B.: Functional dependence. Transactions of the American Mathematical
  Society  \textbf{38}(2),  379--394 (1935)

\bibitem{chen1973invariant}
Chen, B.Y.: An invariant of conformal mappings. Proceedings of the American
  Mathematical Society  \textbf{40}(2),  563--564 (1973)

\bibitem{corman2017functional}
Corman, E., Solomon, J., Ben-Chen, M., Guibas, L., Ovsjanikov, M.: Functional
  characterization of intrinsic and extrinsic geometry. ACM Transactions on
  Graphics (TOG)  \textbf{36}(2), ~14 (2017)

\bibitem{crane2011spin}
Crane, K., Pinkall, U., Schr{\"o}der, P.: Spin transformations of discrete
  surfaces. ACM Transactions on Graphics (TOG)  \textbf{30}(4), ~104 (2011)

\bibitem{farkas1992riemann}
Farkas, H.M., Kra, I.: Riemann surfaces. In: Riemann surfaces, pp. 9--31.
  Springer (1992)

\bibitem{gehring1992topics}
Gehring, F.: Topics in quasiconformal mappings. In: Quasiconformal Space
  Mappings, pp. 20--38. Springer (1992)

\bibitem{gu2003computing}
Gu, X., Wang, Y., Yau, S.T., et~al.: Computing conformal invariants: Period
  matrices. Communications in Information \& Systems  \textbf{3}(3),  153--170
  (2003)

\bibitem{gu2003surface}
Gu, X., Yau, S.T.: Surface classification using conformal structures. In: null.
  p.~701. IEEE (2003)

\bibitem{haantjes1937conformal}
Haantjes, J.: Conformal Representation of an N-dimensional Euclidean Space with
  a Non-definite Fundamental Form on Itself (1937)

\bibitem{Hu2011AClassofIsometric}
Hu, P.: A Class of Isometric Invariants and Their Applications(in Chinese).
  Ph.D. thesis, Institute of Computing Technology, Chinese Academy of Sciences
  (May 2011)

\bibitem{hu2009construction}
Hu, P., Li, H., Lin, Z.: A construction method for surface isometric
  invariants. Journal of Systems Science and Mathematical Sciences  \textbf{9},
  ~006 (2009)

\bibitem{kharevych2006discrete}
Kharevych, L., Springborn, B., Schr{\"o}der, P.: Discrete conformal mappings
  via circle patterns. ACM Transactions on Graphics (TOG)  \textbf{25}(2),
  412--438 (2006)

\bibitem{kuhnel2007liouville}
K{\"u}hnel, W., Rademacher, H.B.: Liouville's theorem in conformal geometry.
  Journal de math{\'e}matiques pures et appliqu{\'e}es  \textbf{88}(3),
  251--260 (2007)

\bibitem{lax1999change}
Lax, P.D.: Change of variables in multiple integrals. The American mathematical
  monthly  \textbf{106}(6),  497--501 (1999)

\bibitem{li2017shape}
Li, E., Huang, Y., Xu, D., Li, H.: Shape dna: Basic generating functions for
  geometric moment invariants. arXiv preprint arXiv:1703.02242  (2017)

\bibitem{li2017isomorphism}
Li, E., Li, H.: Isomorphism between differential and moment invariants under
  affine transform. arXiv preprint arXiv:1705.08264  (2017)

\bibitem{li2018image}
Li, E., Mo, H., Xu, D., Li, H.: Image projective invariants. IEEE Transactions
  on Pattern Analysis and Machine Intelligence  (2018)

\bibitem{liouville1850extension}
Liouville, J.: Extension au cas des trois dimensions de la question du
  trac{\'e} g{\'e}ographique. Applications de l’analyse {\`a} la
  g{\'e}om{\'e}trie pp. 609--617 (1850)

\bibitem{luo2004combinatorial}
Luo, F.: Combinatorial yamabe flow on surfaces. Communications in Contemporary
  Mathematics  \textbf{6}(05),  765--780 (2004)

\bibitem{mullen2008spectral}
Mullen, P., Tong, Y., Alliez, P., Desbrun, M.: Spectral conformal
  parameterization. Computer Graphics Forum  \textbf{27}(5),  1487--1494 (2008)

\bibitem{olver1995equivalence}
Olver, P.J.: Equivalence, invariants and symmetry. Cambridge University Press
  (1995)

\bibitem{ovsjanikov2012functional}
Ovsjanikov, M., Ben-Chen, M., Solomon, J., Butscher, A., Guibas, L.: Functional
  maps: a flexible representation of maps between shapes. ACM Transactions on
  Graphics (TOG)  \textbf{31}(4), ~30 (2012)

\bibitem{rasila2006introduction}
Rasila, A.: Introduction to quasiconformal mappings in n-space. Proceedings of
  the International Workshop on Quasiconformal  (2006)

\bibitem{rustamov2013map}
Rustamov, R.M., Ovsjanikov, M., Azencot, O., Ben-Chen, M., Chazal, F., Guibas,
  L.: Map-based exploration of intrinsic shape differences and variability. ACM
  Transactions on Graphics (TOG)  \textbf{32}(4), ~72 (2013)

\bibitem{springborn2008conformal}
Springborn, B., Schr{\"o}der, P., Pinkall, U.: Conformal equivalence of
  triangle meshes. ACM Transactions on Graphics (TOG)  \textbf{27}(3), ~77
  (2008)

\bibitem{vaxman2015conformal}
Vaxman, A., M{\"u}ller, C., Weber, O.: Conformal mesh deformations with
  m{\"o}bius transformations. ACM Transactions on Graphics (TOG)
  \textbf{34}(4), ~55 (2015)

\bibitem{wang2013affine}
Wang, Y., Wang, X., Zhang, B.: Affine differential invariants of functions on
  the plane. Journal of Applied Mathematics  \textbf{2013} (2013)

\bibitem{white1973global}
White, J.H.: A global invariant of conformal mappings in space. Proceedings of
  the American Mathematical Society  \textbf{38}(1),  162--164 (1973)

\bibitem{willmore2000surfaces}
Willmore, T.J.: Surfaces in conformal geometry. Annals of Global Analysis and
  Geometry  \textbf{18}(3-4),  255--264 (2000)

\bibitem{xu2008geometric}
Xu, D., Li, H.: Geometric moment invariants. Pattern recognition
  \textbf{41}(1),  240--249 (2008)

\bibitem{xu2018content}
Xu, J., Kang, H., Chen, F.: Content-aware image resizing using quasi-conformal
  mapping. The Visual Computer  \textbf{34}(3),  431--442 (2018)

\bibitem{yu2017intrinsic}
Yu, X., Lei, N., Wang, Y., Gu, X.: Intrinsic 3d dynamic surface tracking based
  on dynamic ricci flow and teichm{\"u}ller map. Proceedings. IEEE
  International Conference on Computer Vision  \textbf{2017},  5400--5408
  (2017)

\bibitem{zeng2011registration}
Zeng, W., Gu, X.D.: Registration for 3d surfaces with large deformations using
  quasi-conformal curvature flow. Computer Vision and Pattern Recognition
  (CVPR)  (2011)

\bibitem{zhang2017fast}
Zhang, H., Mo, H., Hao, Y., Li, S., Li, H.: Fast and efficient calculations of
  structural invariants of chirality. arXiv preprint arXiv:1711.05866  (2017)

\end{thebibliography}

\end{document}